\documentclass[twocolumn,prb,showpacs,multicol,amsmath,amssymb]{revtex4}
\usepackage[dvips]{graphicx}
\usepackage{graphicx}
\usepackage{dcolumn}
\usepackage{bm}
\usepackage{graphics}
\usepackage{epsfig,color}
\usepackage{float}

\newcommand{\be}{\begin{equation}}
\newcommand{\ee}{\end{equation}}
\newcommand{\bea}{\begin{eqnarray}}
\newcommand{\eea}{\end{eqnarray}}
\newcommand{\la}{\langle}
\newcommand{\ra}{\rangle}

\begin{document}

\title{ Quantum correlations in the spin-1/2 Heisenberg XXZ chain with modulated Dzyaloshinskii-Moriya interaction}

\author{F. Khastehdel Fumani$^{1}$, B. Beradze$^{2,3}$, 
S. Nemati$^{4,5}$,\\ S. Mahdavifar$^{1}$ and G. I. Japaridze$^{2,3}$}\vspace{2mm}
\affiliation{
$^{1}$ Department of Physics, University of Guilan, 41335-1914, Rasht, Iran\\
$^{2}$  Ilia State University, Faculty of Natural Sciences and Medicine, 0162, Tbilisi, Georgia \\
$^{3}$ Andronikashvili Institute of Physics, 0177, Tbilisi, Georgia\\
$^{4}$ Center of Physics of University of Minho and University of Porto, P-4169-007 Oporto, Portugal\\
$^{5}$ University of Aveiro, Department of Physics, P-3810193 Aveiro, Portugal}


\begin{abstract}

We study a one-dimensional spin-1/2 XXZ Heisenberg model with alternating
Dzyaloshinskii-Moriya interaction, using the numerical Lanczos method. Recently, the ground state (GS) phase diagram of this model has been established 
using the bosonization approach and extensive density matrix renormalization group computations. Four quantum phases -- saturated ferromagnetic (FM), Luttinger liquid (LL), and  two (C1 and C2) gapped phases with composite structure of GS order, characterized by the coexistence of long-range alternating dimer, chirality and antiferromagnetic order have been identified. Here we reexamine the same problem using the exact diagonalization Lanczos method for 
chains up to $N=26$ sites and explicitly detect positions of quantum critical points (QCP) by investigating the quantum correlations as the entanglement and the quantum discord (QD). It is shown that the entanglement quantified by concurrence and the first derivative of the QD are able to reveal besides the standard FM QCP also the Berezinskii-Kosterlitz-Thouless (BKT) phase transition point between the LL and the gapped C1 phase and the Ising type critical point separating the C1 and C2 phases.

\end{abstract}
\maketitle


\section{Introduction}\label{sec1}

The study of quantum phase transitions (QPT) has emerged as an intensely developing field of research in contemporary condensed matter physics \cite{1,2,3,4}.  
Although the correlated electron systems supply a rich variety of models that may be driven to a quantum phase transition \cite{3,5,6}, the low-dimensional quantum magnets remain the most natural prototype systems to study these fascinating  phenomena  \cite{2,4,7,8,9}. 

Quantum phase transitions occur at zero temperature where no thermal fluctuations are present and result from the change of character and the intensity of quantum fluctuations. At absolute zero temperature, with proper tuning of the corresponding model parameter, one can tilt the subtle balance between the competing GSs associated with conflicting interactions of quantum nature. In analogy  with the classical phase transitions, specific values of the model parameters, corresponding to the quantum phase transitions are called the quantum critical points (QCP).
 Identification of the QCP in quantum many-body model is a complicated task and is the subject of keen interest of researchers  over decades.

The one-dimensional spin $S=1/2$ XXZ Heisenberg model with nearest-neighbour interaction 
\begin{eqnarray}
H&=&J\sum_{n}\big[\,S^{x}_{n}S^{x}_{n+1}+S^{y}_{n}S^{y}_{n+1} +\gamma S^{z}_{n}S^{z}_{n+1}\,\big]\, ,
\label{Hamiltonian_XXZ_Heisenberg}
\end{eqnarray}
where $S^{i}_{n}$ ($i=x,y,z$) is the spin-1/2 operator on the $n$-th site and $J>0$, is characterized by the rich GS properties and therefore proved to be an excellent prototype model for studying quantum critical points. Integrability of  the Eq. (\ref{Hamiltonian_XXZ_Heisenberg}) together with the continuum-limit bosonization analysis allows to identify the exact positions of all QCPs and to visualize the effect of the quantum fluctuations via the infrared (large distance)  behavior of spin-spin correlation functions in different phases (for details see \cite{10,11}). Depending on the value of the exchange anisotropy parameter $\gamma$ following three quantum phases are realized in the GS of the model: the ferromagnetic phase (FM)  at $\gamma<-1$, the N\'{e}el phase (AFM) with staggered magnetization reduced by quantum fluctuations at  $\gamma > 1$ and the gapless Luttinger-liquid (LL) phase at $-1<\gamma \leq 1$ characterized with power-law decay of correlations at large distances.

In the recent publication \cite{12}, the ground state phase diagram of the extended version of the one dimensional (1D) spin $S=1/2$ XXZ Heisenberg model 
given by the Hamiltonian 
\begin{eqnarray}
H&=&J\sum_{n}\Big[\, S^{x}_{n}S^{x}_{n+1}+S^{y}_{n}S^{y}_{n+1} +\gamma S^{z}_{n}S^{z}_{n+1}\nonumber\\
&+&i(d_{0}+(-1)^{n}d_{1})\left(S^{x}_{n}S^{y}_{n+1}-S^{y}_{n}S^{x}_{n+1}\right) \Big],
\label{Hamiltonian_Alt_DM_XXZ_Heisenberg}
\end{eqnarray}
has been studied using the bosonization approach and extensive density matrix renormalization group computations. The second term in Eq.(\ref{Hamiltonian_Alt_DM_XXZ_Heisenberg}) corresponds to the Dzyaloshinskii-Moriya (DM) interaction 
\begin{eqnarray}
H_{DM}&=&J\sum_{n}{\bf d}(n)\cdot[{\bf S}_{n}\times{\bf S}_{n+1}],
\label{Hamiltonian_DMI}
\end{eqnarray}
in the case of orientated along the \^{z} axis alternating axial vector 
\begin{equation}
{\bf d}(n)=\left(0,0, d_{0}+(-1)^{n}d_{1}\right)\, .
\end{equation}

The Dzyaloshinskii-Moriya interaction corresponds to the antisymmetric part of the exchange tensor and appears in a system with broken inversion symmetry due to the spin-orbit coupling. Historically, it was first introduced by I. Dzyaloshinskii on the grounds of general symmetry arguments to explain weak canted ferromagnetism observed in hematite $\alpha-Fe_2O_3$ \cite{13}. Later, the spin-orbit coupling as the microscopic mechanism of the antisymmetric exchange interaction has been identified by T. Moriya \cite{14}. 

The intensity of the DM coupling between two neighboring spins is determined by the vector ${\bf d}$, which usually is established using the Moriya symmetry rules \cite{14} or, in more advanced cases,  within the L\'{e}vy and Fert model \cite{15} or by the first--principles calculations \cite{16,17}. Generally, in a spin chain, vector ${\bf d}$ may vary on neighboring links both 
in magnitude and orientation, however,  the symmetry  restrictions based on the properties of real solid state materials usually rule out most of the possibilities and confine  the majority of theoretical discussion to two principal cases -- uniform DM interaction, ${\bf d}$ vector remains unchanged 
over the system \cite{18,19,20,21,22,23,24,25,26} and the case of staggered DM interaction, with the antiparallel orientation of ${\bf d}$ on adjacent bonds \cite{27,28,29,30,31,32,33,34,35,36,37,38,39,40,41,42,43}. 
In both cases, the DM term can be eliminated by the gauge transformation and absorbed in a boundary condition not affecting the basic structure
of the Hamiltonian \cite{18,19}. Therefore, acting separately, neither uniform nor a staggered DM term can change the spectral properties of the 1D spin chain and trigger the unconventional mechanisms for gap formation only in the presence of applied magnetic field and/or the 
staggered gyromagnetic tensor. Real systems that show this type of behavior are those which display chain structure, such as Cu-benzoate 
\cite{27,28,29,30}, Cu-Pyrimidine Dinitrate \cite{31,32,33}, $Yb_4As_3$ \cite{34,35,36,37}, $BaCu_2Si_2O_7$ \cite{38,39}, $Sr_2V_3O_9$ \cite{40} as well as the ladder-type structures such as $(C_7H_{10}N)_2CuBr_4$  and $Cu(C_8H_6N_2)Cl_2$ \cite{44,45,46}. 

Recently it has been demonstrated that the DM interaction can be efficiently tailored with a substantial efficiency factor by the  external electric field \cite{47,48,49}. This unveils the possibility to control DM interaction and magnetic anisotropy via the electric field and opens a wide area for application of effects caused by the spatially modulated DM interaction on the properties of low-dimensional spin systems. Relevant recent theoretical prediction applies to the gap formation mechanism in the XXZ Heisenberg chain with modulated DM interaction due to the interplay between the uniform and staggered parts of the DM coupling at  the isotropic $g$-factor and in the absence of an external magnetic field \cite{12}.

\begin{figure}[t]
\centerline{\psfig{file=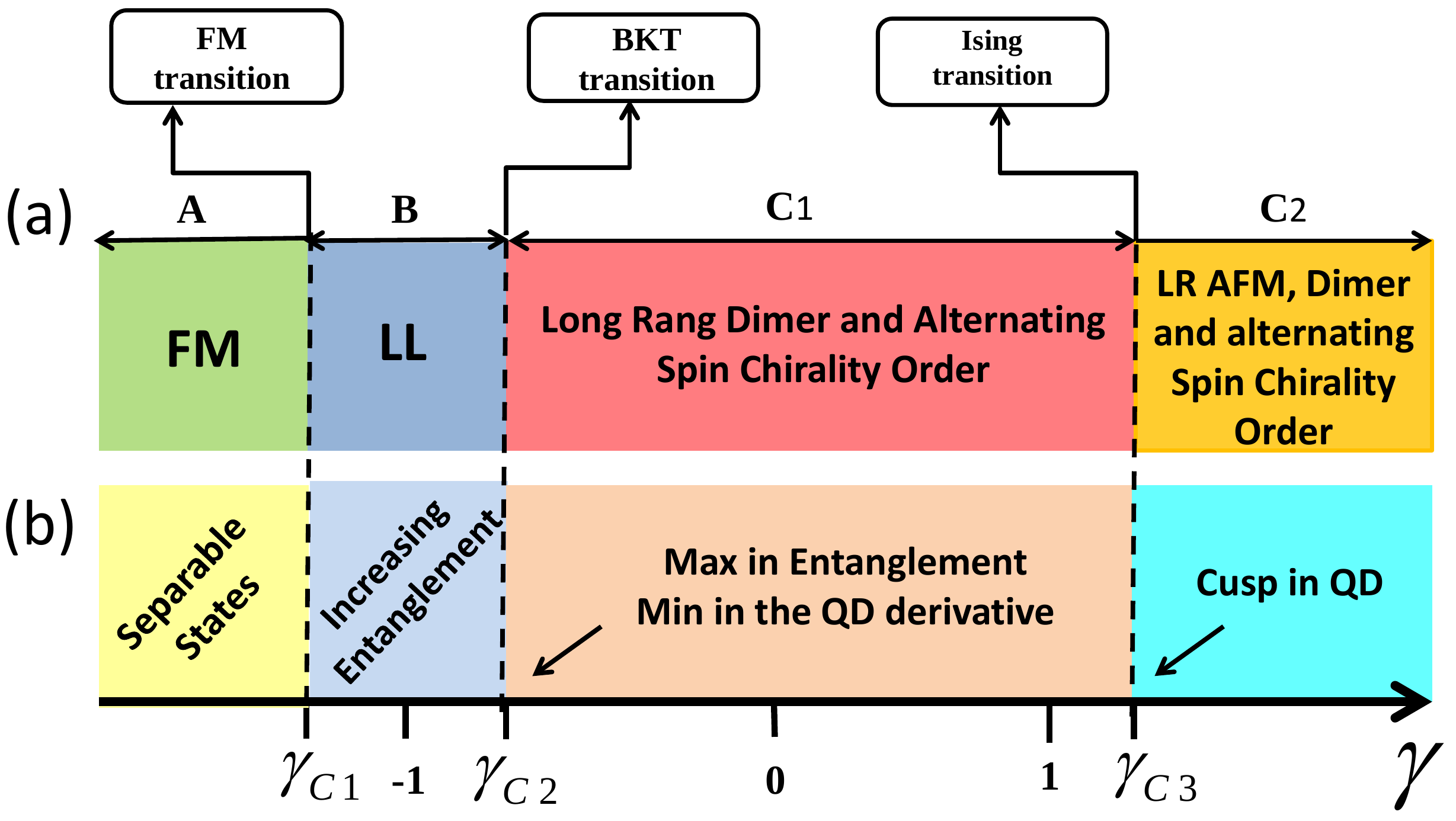,width=3.4in}}
\caption{ (a) Schematic picture of the ground state phase diagram of the spin-$1/2$ XXZ Heisenberg chain with the modulated DM interaction \cite{12}. (b) The same picture from the view point of quantum correlations. 
 }\label{Fig1}
\end{figure}

For $d_0d_1 \neq 0$, depending on the value of the exchange anisotropy parameter $\gamma$, the GS phase diagram of the model (\ref{Hamiltonian_Alt_DM_XXZ_Heisenberg}) contains three QCP, which separate the following four different phases (see Fig.~\ref{Fig1}~(a)). 

The first quantum critical point, with the exact value 
\begin{equation}
\gamma_{c1}=-J^{\ast}=- \sqrt{1+d_{0}^{2}+d_{1}^{2}}~,
\end{equation}
corresponds to the transition from the ferromagnetically ordered phase
for
  $\gamma<\gamma_{c1}$ into the LL phase.
  The interaction coupling constant is considered to be $J=1$. 
The gapless LL phase is shrunk up to a narrow region between $\gamma_{c1}<\gamma\leq\gamma_{c2}=-J^{\ast}/\sqrt{2}$. At $\gamma=\gamma_{c2}$ the Berezinskii-Kosterlitz-Thouless (BKT) phase transition takes the system into the composite ($C1$) gapped phase characterized by the
coexistence of long-range ordered (LRO) spin dimerization
\begin{equation}
\epsilon = \frac{1}{N}\sum_{n=1}^{N}(-1)^{n}\la{\bf S}_{n}\cdot {\bf S}_{n+1}\ra~,
\end{equation}
and alternating spin chirality
\begin{equation}
\kappa = \frac{1}{N}\sum_{n=1}^{N}(-1)^{n}\la \left[\,{\bf S}_{n}\times {\bf
S}_{n+1}\,\right]_{z}\ra ~, 
\end{equation}
patterns. Here the brackets symbolize the GS average values.

Finally, at 
\begin{equation}
\gamma=\gamma_{c3} \simeq J^{\ast} \left[1+\left(d_{0}d_{1}/J^{\ast\,2}\right)^{0.4}\right]~,
\end{equation}
the Ising type phase transition into the other gapped  composite $C2$ phase takes place. In this phase, besides the 
long-range dimerization and alternating spin-chirality order, the long-range antiferromagnetic order, characterized by the staggered 
magnetization 
\begin{equation}
m=\frac{1}{N}\sum_{n=1}^{N}(-1)^{n} \la S_{n}^{z} \ra ~,
\end{equation}
is present. 

In the last years, very powerful approaches based on the implementation of concepts partly borrowed from the quantum information theory \cite{50} have been developed and intensively used to identify QCP in the various models of strongly correlated electron or spin systems \cite{51}. In particular, the detailed analysis of  various bipartite quantum correlations as the entanglement and the quantum discord (QD), calculated in the case of finite clusters, has been successfully used to solve these problems \cite{52,53,54,55,56,57,58,59,60,61,62,63,64,65,66,67,68}. 

In this paper, we use the numerical exact diagonalization Lanczos method for chains up to $N=26$ sites and calculate various bipartite quantum correlations 
as a function of the exchange anisotropy parameter $\gamma$ to determine precise positions of all QCPs of the spin chain model  (\ref{Hamiltonian_Alt_DM_XXZ_Heisenberg}) (see Fig.~\ref{Fig1}~(b)). In the last years this technique has been successfully used to study the GS phase diagram of the Heisenberg \cite{69,70} and Ising spin chains \cite{71,72,73,74,75,76,77}.

The outline of this paper is as follows. In the forthcoming
 Section, 
 we introduce quantum correlations including the concurrence and the QD. In Sec.~\ref{sec3}, we present our exact calculation results of the various two-spin quantum correlations. Finally, the results are discussed and summarized in Sec.~\ref{sec4}.  

\section{Quantum correlations}\label{sec2} 

It is well established, that in the marked contrast with local order parameters used in the case of phase transitions described within the 
Landau-Ginzburg-Wilson paradigm, in quantum systems to trace the signature of quantum criticality the correlations that have no classical analogue,
such as quantum entanglement, have to be considered \cite{50}.
 To quantify entanglement, different methods have been suggested. One can use a measure of the two-party entanglement to determine the shared entanglement  by two particles in a lattice.

In the following, we will use the concurrence, a monotonic function ranging from zero (a separable state) to one (a maximally entangled state), as a measure of the entanglement between a pair of spins \cite{78}. The concurrence of two spin-1/2 particles can be calculated from the 
two-site reduced 
density matrix of the model -- $\rho_{n,m}$. In the standard basis $\{ |\uparrow\uparrow\rangle, |\uparrow\downarrow\rangle, |\downarrow\uparrow\rangle, |\downarrow\downarrow\rangle \}$, the corresponding reduced density matrix is expressed as 
\begin{eqnarray}
\rho_{n,m}=\left(
\begin{array}{cccc}
X_{nm}^{+} & 0 & 0 & F_{nm}^{*} \\
0 & Y_{nm}^{+} & Z_{nm}^{*} & 0 \\
0 & Z_{nm} & Y_{nm}^{-} & 0 \\
F_{nm} & 0 & 0 & X_{nm}^{-} \\
\end{array}
\right),
\label{density matrix2}
\end{eqnarray}
where its non-zero elements are defined as 
\begin{eqnarray} 
X_{nm}^{+}&=& \langle(1/2+S_{n}^{z})(1/2+S_{m}^{z})\rangle,\nonumber\\
Y_{nm}^{+}&=& \langle(1/2+S_{n}^{z})(1/2-S_{m}^{z})\rangle,\nonumber\\
Y_{nm}^{-}&=& \langle(1/2-S_{n}^{z})(1/2+S_{m}^{z})\rangle,\\
X_{nm}^{-}&=& \langle(1/2-S_{n}^{z})(1/2-S_{m}^{z})\rangle,\nonumber\\
Z_{nm}&=& \langle S_n^{+}S_{m}^{-}\rangle,\nonumber\\
F_{nm}&=& \langle S_n^{+}S_{m}^{+}\rangle.\nonumber
\label{dm2}
\end{eqnarray}
 The concurrence is given by the following expression:
\begin{eqnarray} 
C_{nm} =2 \max{\{0, \Lambda^{(1)}_{nm},\Lambda^{(2)}_{nm}\}}
\label{Concurr}
\end{eqnarray}
where
\begin{eqnarray} 
\Lambda^{(1)}_{nm} &=& |Z_{nm}|-\sqrt{X_{nm}^{+}X_{nm}^{-}}\, ,\nonumber\\ 
\Lambda^{(2)}_{nm} &=& |F_{nm}|-\sqrt{Y_{nm}^{+}Y_{nm}^{-}}\nonumber\, .
\end{eqnarray}

On the other hand,
besides the entanglement which has been defined according to the separability of states, a general quantity known as QD contains all quantum correlations in the considered systems.
Zurek  defined the QD as a measure of 'non-classicality' that can also exist in the separable states \cite{79,80}.

The QD between a pair of spins located at  $n$ and $m$ sites is defined as
\begin{equation}
QD_{nm} = {\cal I}({\rho _{n,m}}) - {\cal C}({\rho _{n,m}}),
\label{QD}
\end{equation}
where $\cal I$ denotes the mutual information and $\cal C$ stands for the classical correlations (for detailed calculations see Ref.~\cite{81}). Although, the QD borrowed from the quantum information theory, it has been proved to be a powerful tool for detecting QCPs in the complicated systems.

\section{Results}\label{sec3}

In what follows we use the results of the numerical exact diagonalization Lanczos method to calculate quantum correlations and 
analyze their behavior throughout the GS phases of the model (\ref{Hamiltonian_Alt_DM_XXZ_Heisenberg}) using the exchange anisotropy $\gamma$
as a scanning parameter. We restrict our calculations to the case $J > 0$ and  take $J=1$ without losing generality.
At the first step, we calculate the GS of the system for different chain sizes, $N=16, 20, 22, 24$ and $26$, then we calculate the density matrix of a pair of spins with respect to the GS of the finite chains of the given size and finally, using the relations
Eqs. (\ref{Concurr}) and (\ref{QD})
 calculate the required quantum correlation functions.
\subsection{\textit{Case of Uniform ($d_0\neq 0$,\, $d_1=0$) and Staggered ($d_0 =0 $, \, $d_{1} \neq 0$) DM interaction}}

As it was shown in Ref.~\cite{12} the most direct way to show up properties of the model (\ref{Hamiltonian_Alt_DM_XXZ_Heisenberg}) is
to gauge away the DM interaction at the very first stage and to rewrite the Hamiltonian in a physically more suggestive form.  Performing the site-dependent 
rotation of spins around the $z$-axis as introducing new spin variables $\tau_n^{x,y,z}$  
\begin{eqnarray} 
S^{x}_{2n-1}&=&\cos[(n-1)(\theta_-+\theta_+)]\,\tau^{x}_{2n-1}\nonumber \\
&-&\sin [(n-1)(\theta_-+\theta_+)]\,\tau^{y}_{2n-1}\, ,\\
S^{y}_{2n-1}&=&\sin[(n-1)(\theta_-+\theta_+)]\,\tau^{x}_{2n-1}\nonumber \\
&+&\cos[(n-1)(\theta_-+\theta_+)]\,\tau^{y}_{2n-1}\, ,  \\
S^{x}_{2n}&=&\cos[n(\theta_-+\theta_+)-\theta_+]\,\tau^{x}_{2n}\nonumber \\
&-&\sin [n(\theta_-+\theta_+)-\theta_+]\,\tau^{y}_{2n}\, ,\\
S^{y}_{2n}&=&\sin [n(\theta_-+\theta_+)-\theta_+]\,\tau^{x}_{2n}\nonumber \\
&+&\cos [n(\theta_-+\theta_+)-\theta_+]\,\tau^{y}_{2n}\, ,
\label{eq27-1}
\end{eqnarray} 
and choosing angles $\theta_{\pm}$ such that $\tan(\theta_{\pm})=d_0 \pm d_1$ and obtain the Hamiltonian without the DM interaction but only with the alternating transverse exchange interaction \cite{82}
\begin{eqnarray}
\label{XXZ_w_Modulated_DMI_Rotated} {\cal H}&=& J^{\ast}
\sum_{n}
\Big[\,(1+(-1)^{n}\delta)\left(\tau^{x}_{n}\tau^{x}_{n+1}
+ \tau^{y}_{n}\tau^{y}_{n+1}\right)\nonumber\\
 &&\hspace{12mm} \,+ \gamma_{eff} \tau^{z}_{n}\tau^{z}_{n+1}\Big]\, ,
\end{eqnarray}
where $J^{\ast}=J\sqrt{1+d^2_0+d^2_1}$, $\gamma_{eff}=\gamma/\sqrt{1+d^2_0+d^2_1}$ and at $d_{i}=d_0 \pm d_1 \ll 1$ ( $i=\pm$),
\begin{eqnarray}
&&\delta\, \simeq d_0d_1+ {\cal O}\left(d_{i}^{4}\right)\, .
\end{eqnarray}

Thus at $\delta=0$ i.e. in the case of  uniform ($d_1 = 0$) or staggered ($d_0=0$) DM interaction, the effective model after
gauging  away the DM interaction is the Hamiltonian of XXZ chain (\ref{Hamiltonian_XXZ_Heisenberg}) with renormalized exchange anisotropy parameter $\gamma_{eff}$.
Comparison with the exact GS phase diagram \cite{10,11} gives that the FM phase is realized at $\gamma <-\sqrt{1+d_{i}^{2}}$, the   gapless chiral liquid  phase at $-\sqrt{1+d_{i}^{2}}<\gamma\leq \sqrt{1+d_{i}^{2}}$ and LRO AFM phase at $\gamma > \sqrt{1+d_{i}^{2}}$,  where 
$i=0$ ($i=1$) in the case of uniform (staggered) DM interaction, respectively.

\begin{figure}
\centerline{\psfig{file=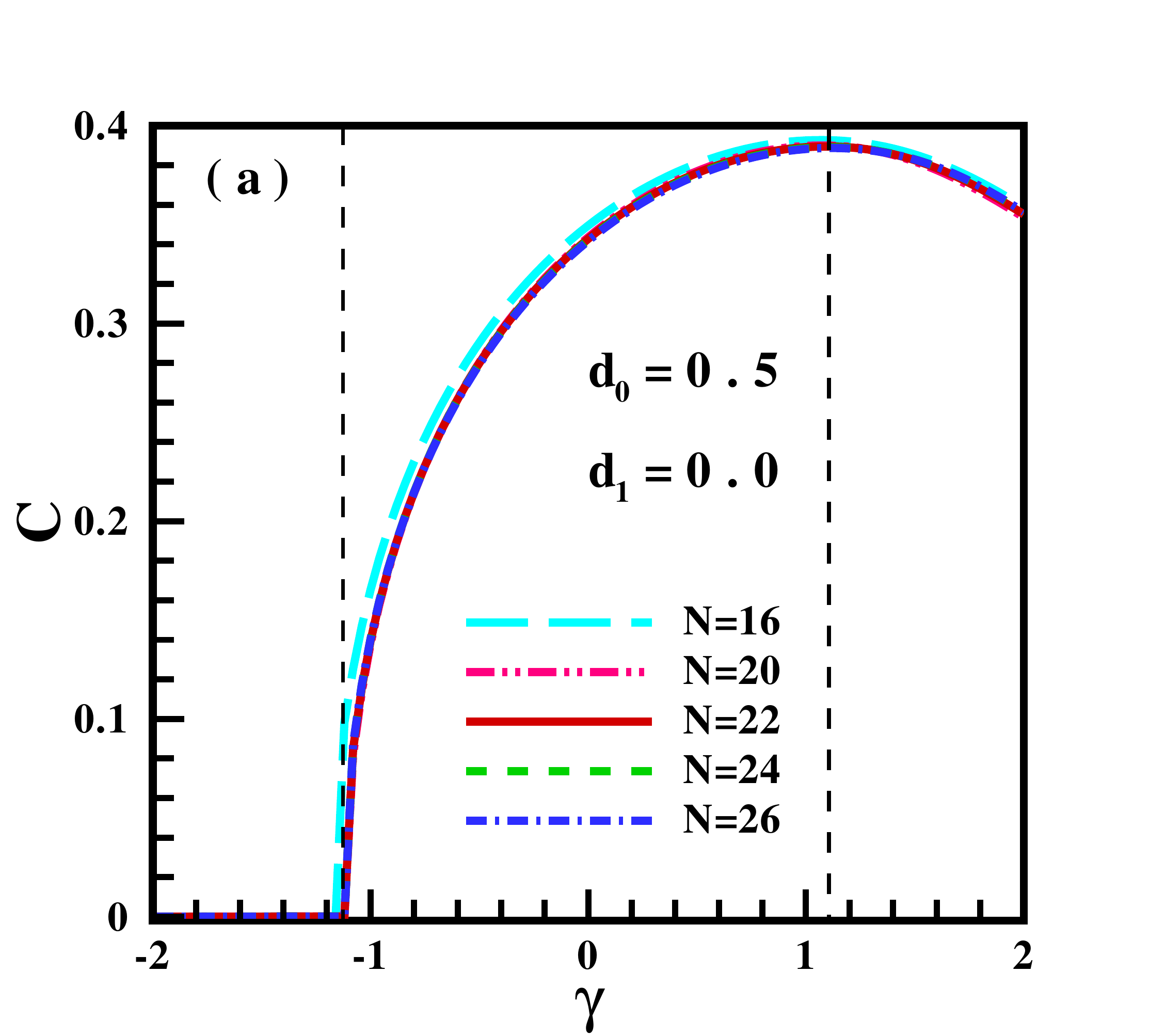,width=1.8in}  \psfig{file=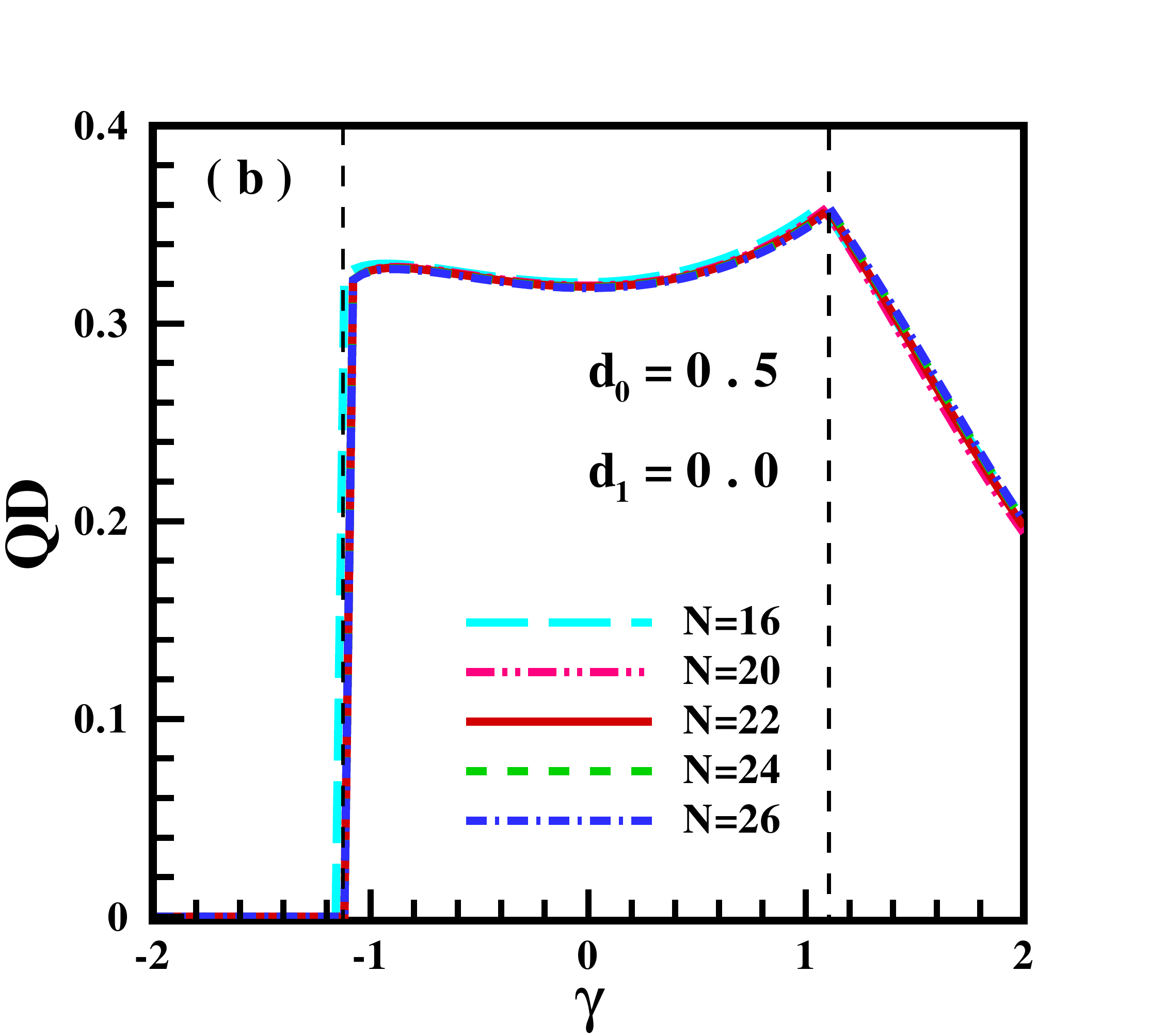,width=1.8in}}
\centerline{\psfig{file=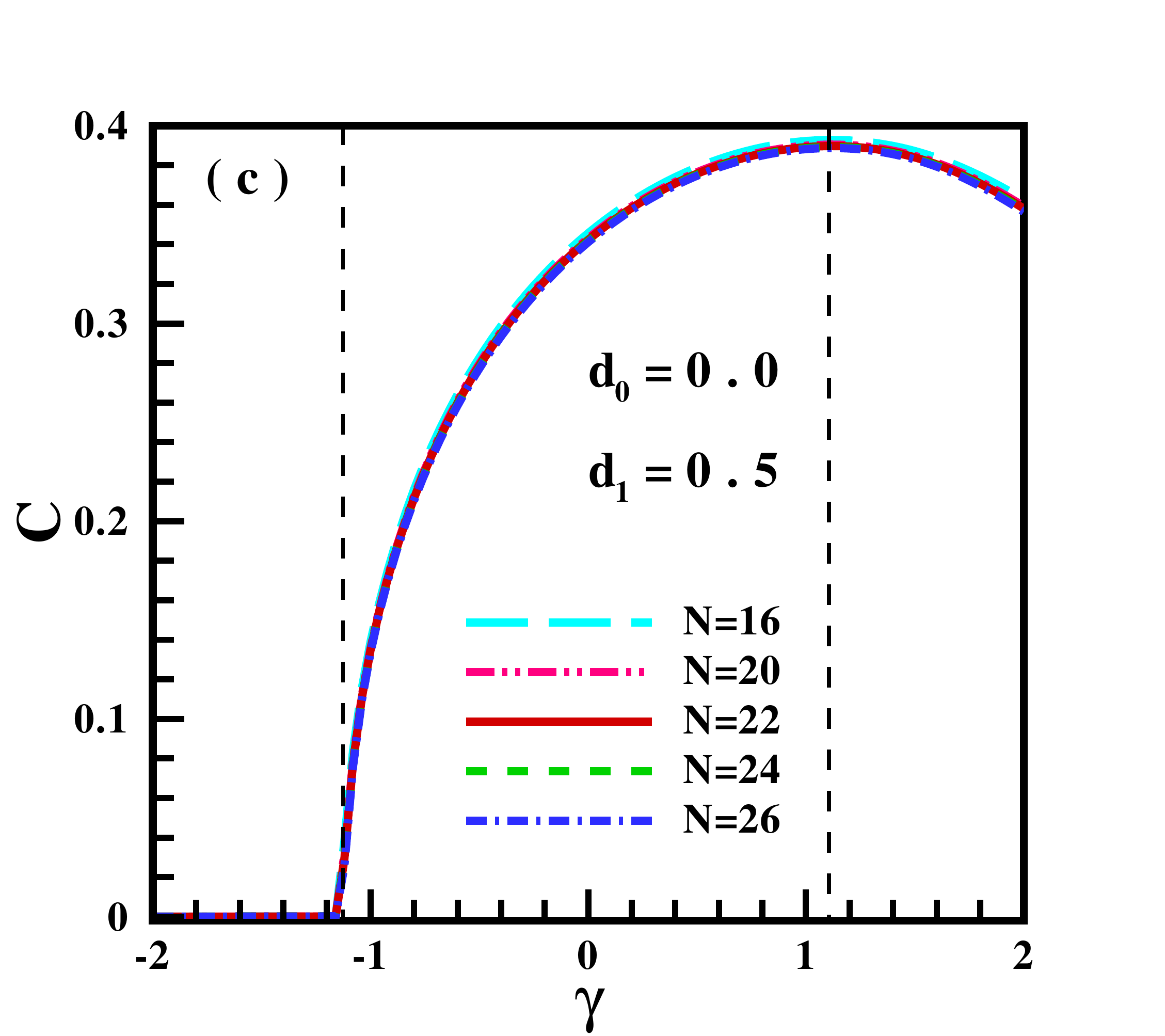,width=1.8in}  \psfig{file=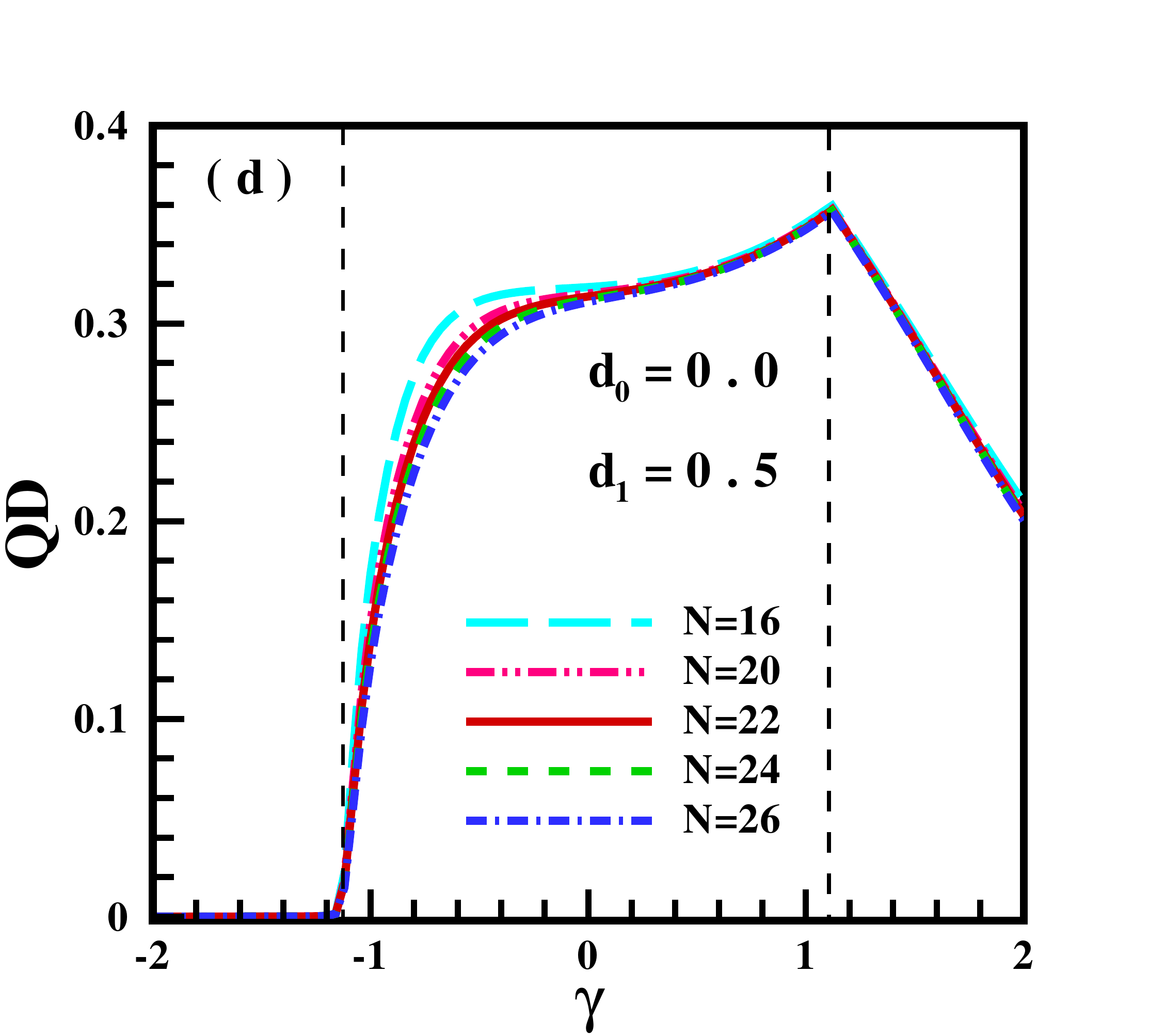,width=1.8in}}
\caption{
  (a) The concurrence (b)  the quantum discord  between the nearest neighbor spin pairs as a function of $\gamma$ 
for $d_0=0.5$, $d_1=0$  (c) the concurrence, and (d)  the quantum discord  between the nearest neighbor spin pairs as a function of $\gamma$ for $d_0=0$, $d_1=0.5$, using Lanczos method for chain sizes $N=16, 20, 22, 24,$ and $26$ calculated at $J=1$.
}
\label{D1=0}
\end{figure}

In Figs.~\ref{D1=0}~(a) and (b), the concurrence and the QD, as a function of $\gamma$ at $J=1$ and $d_0=0.5$ and $d_1=0.0$ are plotted, calculated in the case of 
chains of length $N=16, 20, 22, 24$, and $26$. According to the panel (a), in the saturated FM phase, the concurrence of nearest pair spins is zero which is in complete agreement  with the fact that the saturated FM state is a separable quantum state. Exactly at the first critical point, $\gamma=- \sqrt{1+d_0^2}= -1.11 \pm 0.01$, concurrence starts to increase and gains its maximum amount at  $\gamma= \sqrt{1+d_0^2}= 1.11 \pm 0.01$ where a transition from the chiral phase into the N\'{e}el phase occurs.

Furthermore, the nearest pair spins stay entangled in the N\'{e}el phase, although the entanglement value decreases by increasing $\gamma$ in this phase.  
The results of computed QD between the nearest pair spins is shown in the panel (b) of Fig. \ref{D1=0}. 
Similar to the concurrence, the QD has zero value in the saturated FM phase.
 By increasing $\gamma$, one can explicitly identify two quantum critical points which mark a qualitative change 
 in the behavior of QD as a function of $\gamma$ without display of any valuable finite size effect. The cusp form at the second critical point denotes non-analytic behavior of the QD known as an indication of the second order quantum phase transition \cite{61,83,84}. Our results are in a
complete agreement  with  the results of  previous studies on the pure spin-1/2 XXZ chain model \cite{85,86}.

In Figs.~\ref{D1=0}~(c) and (d), results of the similar exact diagonalization calculations of the concurrence and of the QD,  at $J=1$ and $d_0=0$ and $d_1=0.5$ and $N=16, 20, 22, 24$, and $26$, are presented. Comparison of these data with corresponding results obtained in the case of uniform DM show the concurrence values are the same in two different cases $d_1=0$ or $d_0=0$ (provided by the same value of the parameter $J^{\ast}$) being an additional 
proof of the universality of the effective Hamiltonian
Eq. (\ref{XXZ_w_Modulated_DMI_Rotated}),
  obtained by the gauge transformation in both the above considered cases. On the other hand, the QD between the nearest pair spins shows weakly different behavior in the proximity of the first quantum critical point. This can be treated as a sign of the difference between the chiral and the staggered chiral phases from the viewpoint of the quantum correlations.

\subsection{\textit{Modulated DM interaction ($d_0\neq 0$, $d_1 \neq 0$)}}

\begin{figure}
\centerline{\psfig{file=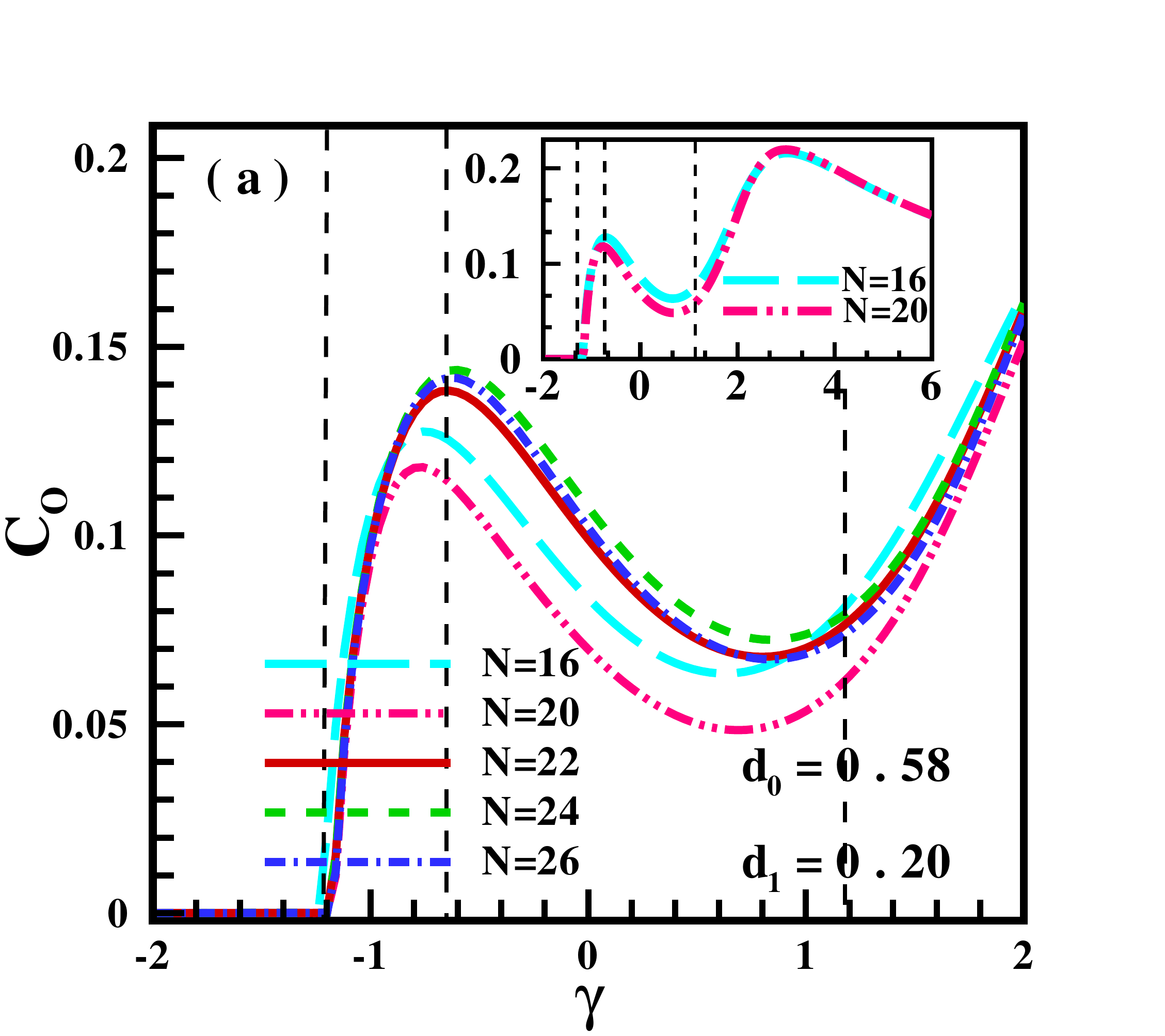,width=1.8in}   \psfig{file=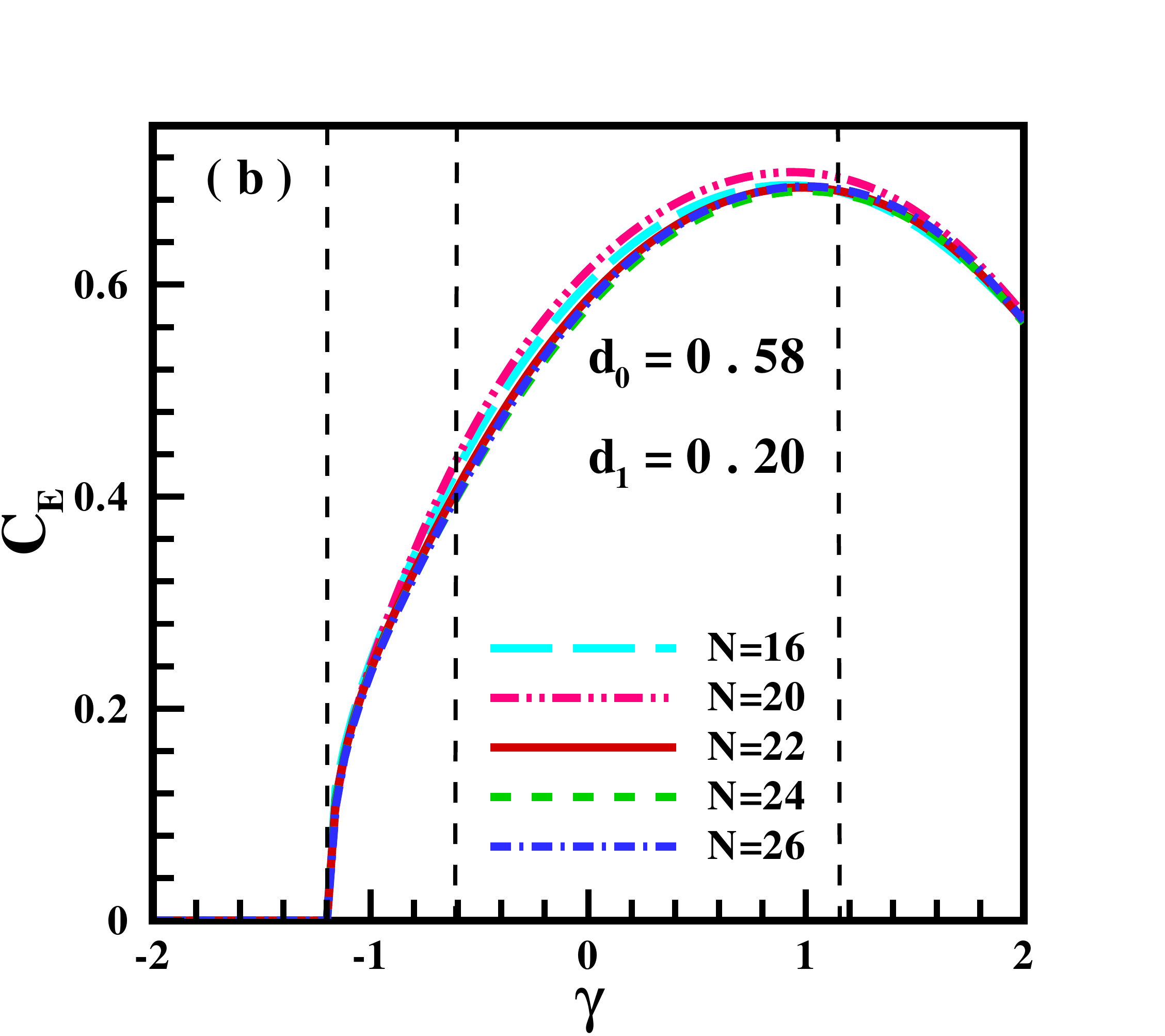,width=1.8in}}    
\centerline{\psfig{file=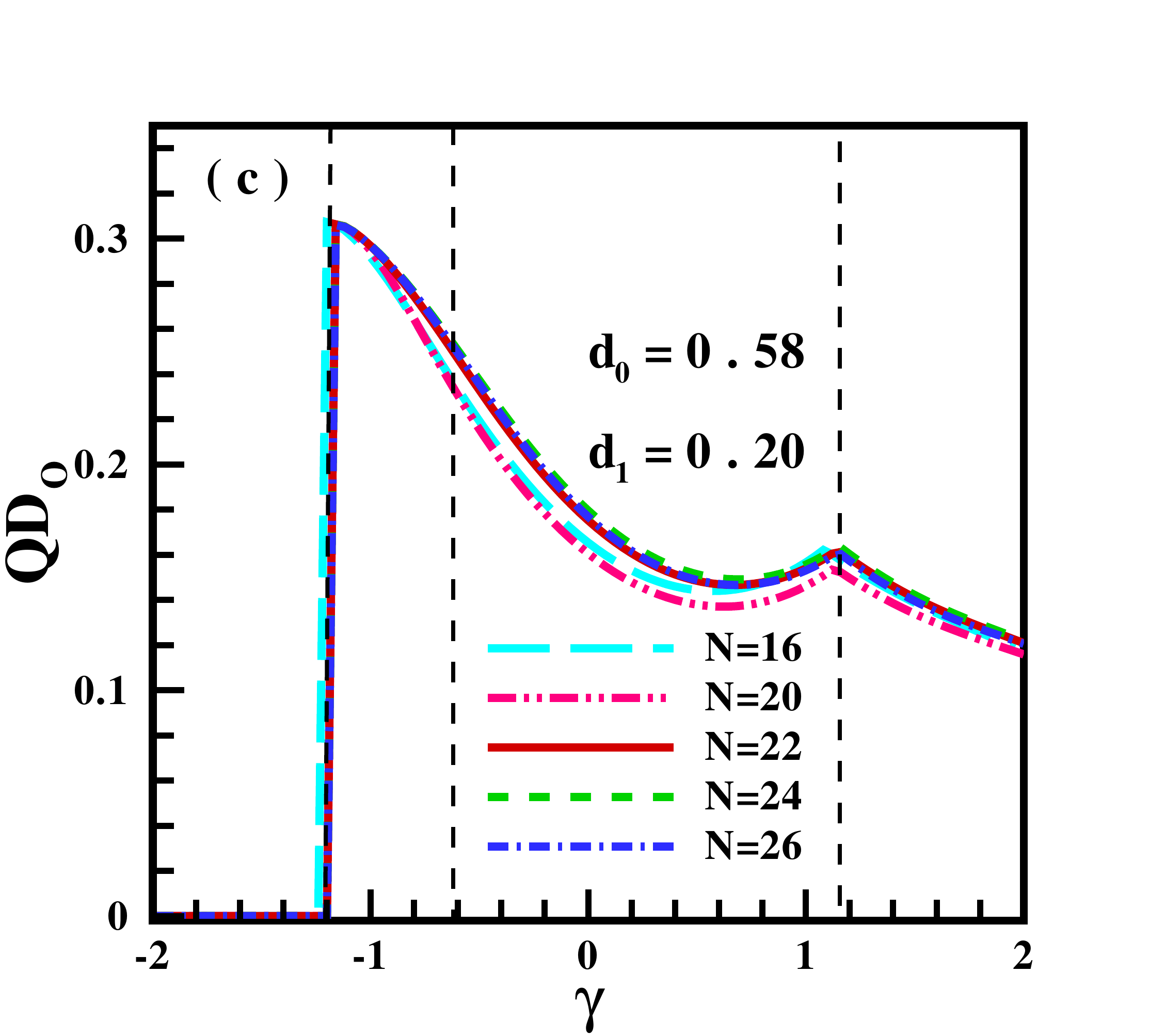,width=1.8in}   \psfig{file=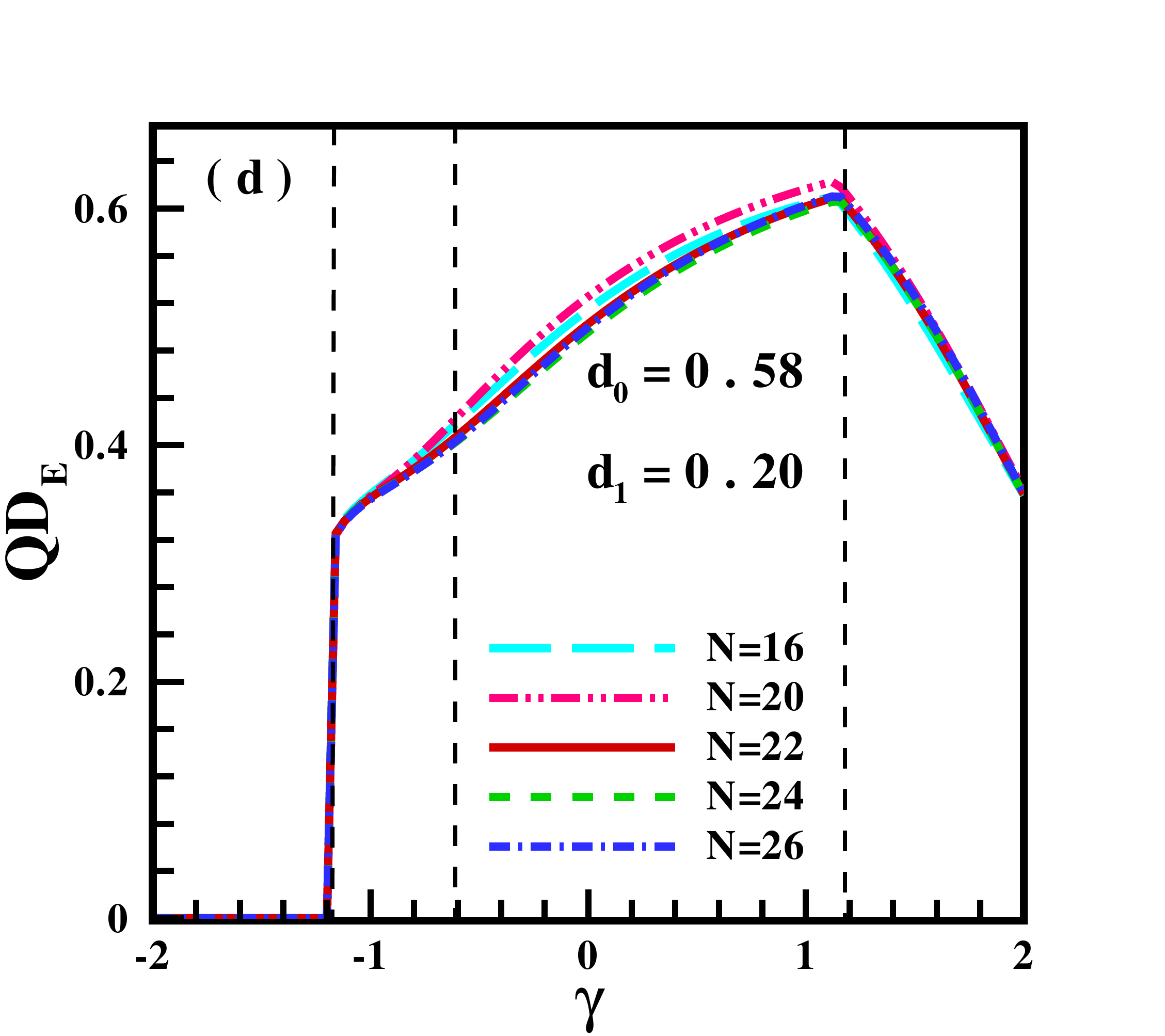,width=1.8in}}   
\caption{ The concurrence between the nearest neighbor spin pairs separated by (a) odd links and (b) even links  and quantum discord between the nearest neighbor spin pairs  separated by (c) odd links and (d) even links  as a function of $\gamma$, for $d_0=0.58$, $d_1=0.2$ and $J=1$, using Lanczos results for chain sizes $N=16, 20, 22, 24,$ and $26$.
The inset in panel (a) shows the concurrence with respect to $\gamma$ separated by odd links for chain sizes $N=16$ and $20$.
}
\label{D0=0.58}
\end{figure}

In this part, we extend the study of quantum correlations in the presence of both uniform and staggered parts of DM interaction. We considered different values of DM interaction and numerically diagonalized the Hamiltonian for different finite-size chains. 
Specifically, we present the numerical results of quantum correlations as a function of $\gamma$ for $d_0=0.58$ and $d_1=0.2$ (Fig. \ref{D0=0.58}).
Since, in this case, the DM interaction on odd links differs from the even links,  therefore, quantum correlations have been calculated between the
nearest pair spins on odd and even links, separately.

In Fig. \ref{D0=0.58}~(a) the concurrence between a pair of spins, separated by odd links, $C_{O}$ as a function of the anisotropy parameter $\gamma$ is depicted for  finite chains $N=16, 20, 22, 24,$ and $26$. The same behavior of the entanglement is observed in all finite size chains. As it is seen, spins  are not entangled up to the first critical point  $\gamma_{c1} \simeq -1.17 \pm 0.01 \simeq - J^{\ast} $.
The concurrence becomes finite as soon as $\gamma$ passes the first critical point $\gamma_{c1}$ and then increases with increasing $\gamma$.
 As we observe  $C_{O}$ reaches a maximum value at the BKT phase transition point, i.e.  $\gamma_{c2} \simeq -0.65 \pm 0.01$ in a very good agreement with the prediction of analytical ($\gamma_{c2}=- \frac{J^*}{\sqrt{2}} \simeq -0.83$) and numerical DMRG ($\gamma_{c2} \simeq 0.84 \pm 0.08$) studies  \cite{12}. The concurrence $C_{O}$ also displays a minimum for large value of  $\gamma \simeq  0.8$ in the $C1$ phase, however no trace of the third critical point $\gamma_{c3}$ can be subtracted from these data. 


To proceed, the entanglement between a pair of spins on even links, $C_{E}$, is demonstrated in Fig. \ref{D0=0.58}~(b).
 Similar to the concurrence between spin pairs on odd links, no signature of the second and third critical points, $\gamma_{c2}$ and $\gamma_{c3}$,  is present in the displayed in the $\gamma$ dependence 
of the concurrence between a pair spins separated by even links on $C_{E}$.
 The DM interaction between a pair of  spins on even links ($d_0+d_1$) is stronger than the odd links. 
  It is known that spin pairs with pure DM interaction are completely entangled.  For this reason, the increase of entanglement by raising the DM interaction is naturally expected \cite{87, 88}.

Due to the higher value of the DM interaction between spins separated by even links than by odd links, the entanglement, $C_{E}$, is significantly stronger than $C_{O}$ as it is clearly seen if one compares data presented in panels (a) and (b) of Fig. \ref{D0=0.58}. As we observe in these figures, non-entangled spins in the FM phase, start to entangle at  $\gamma>\gamma_{c1}\simeq  -1.17 \pm 0.01$. With further increase of $\gamma$, the entanglement also increases, however it shows different behavior, when is calculated for spins separated by even or odd links. 
Concurrence for even links, $C_{E}$, monotonically increases with increasing $\gamma$ and reaches its maximum around the third critical point  $\gamma_{c3} \simeq 1.17 \pm 0.01$. Therefore the concurrence between pair of spins on even links does not trace any sign of the BKT transition. However, the concurrence between pair of spins on weak odd links, $C_{O}$, shows a well-pronounced maximum in the vicinity of the second QCP at  $\gamma_{c2} \simeq\gamma \simeq -0.65 \pm 0.01$ then decays, again starts to increase in the vicinity of $C1-C2$ transition point and finally decreases in the AFM phase (shown in the inset of Fig.~\ref{D0=0.58} (a) for chain size $N=16, 20$).


Panels (c) and (d) in Fig.~\ref{D0=0.58} depict QD between pair of spins on odd and even links, respectively.
As shown in Fig.~\ref{D0=0.58}~(c), we observe a discontinuous behavior for $QD_{O}$ along the quantum phase transition from FM to LL phase without any finite size effect.
In both LL and the first composite phases, $QD_{O}$ has decreasing behavior. With further increase of $\gamma$, the Ising type transition 
from $C1$ onto the $C2$ composite phase at  $\gamma_{c3} \simeq 1.17 \pm 0.01$ is clearly signaled by the maximum (cusp form) in the $\gamma$ dependence of $QD_{O}$. We have to mention that this critical value is also in a good agreement with  the numerical DMRG ($\gamma_{c3} \simeq 1.70 \pm 0.08$) result \cite{12}.   

From Fig.~\ref{D0=0.58}~(d), one can clearly detect QCP between FM and LL phases as a jump in $QD_{E}$ value.
 Moreover, the maximum in Fig.~\ref{D0=0.58}~(d) shows the Ising type transition between $C_1$ and $C_2$ composite phases. In addition, as the same as concurrence the amount of QD between pair of spins on even links is larger than odd links and also shows monotonic increasing behavior in the first composite phase $C_1$. 

\begin{figure}
\centerline{\psfig{file=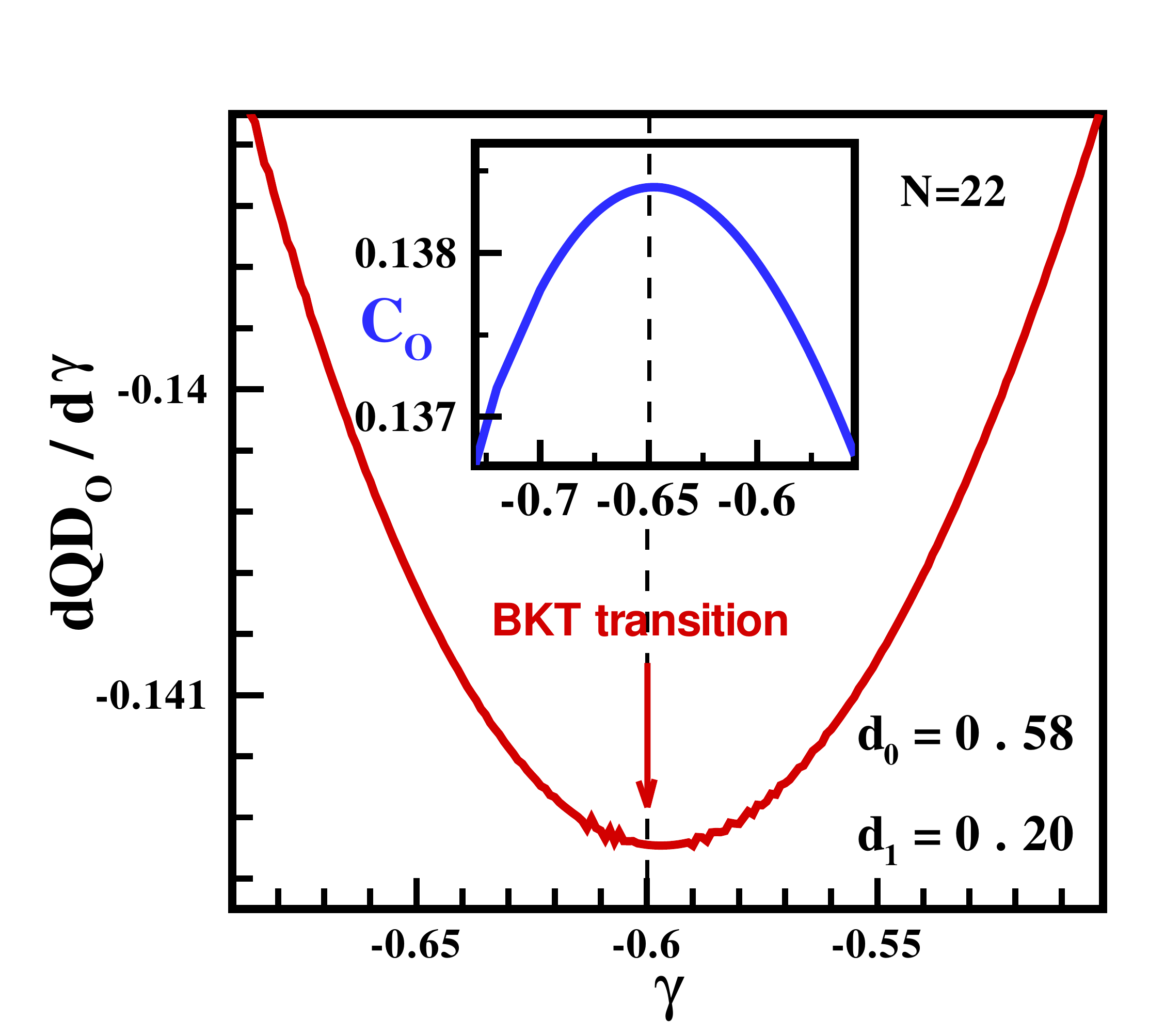,width=2in}}
\caption{ The first derivative of the QD between spin pairs on odd links with respect to the $\gamma$ for a chain size $N=22$. The inset shows the concurrence as a function of $\gamma$ between the mentioned pair of spins.}
\label{D-QD}
\end{figure}

Next, to prob the BKT quantum phase transition point in the derivation of QD,
  we perform the first derivative of $QD_{O}$ for a system size $N=22$ in Fig.~\ref{D-QD}.  As easily seen in Fig.~\ref{D-QD}, the BKT phase transition is clearly signaled by a minimum in the first derivative of the $QD_{O}$. The minimum is 
nearly placed
 where the concurrence between pair of spins on odd links, $C_{O}$, will be maximized (shown in the inset of Fig.~\ref{D-QD}).

\begin{figure}
\centerline{\psfig{file=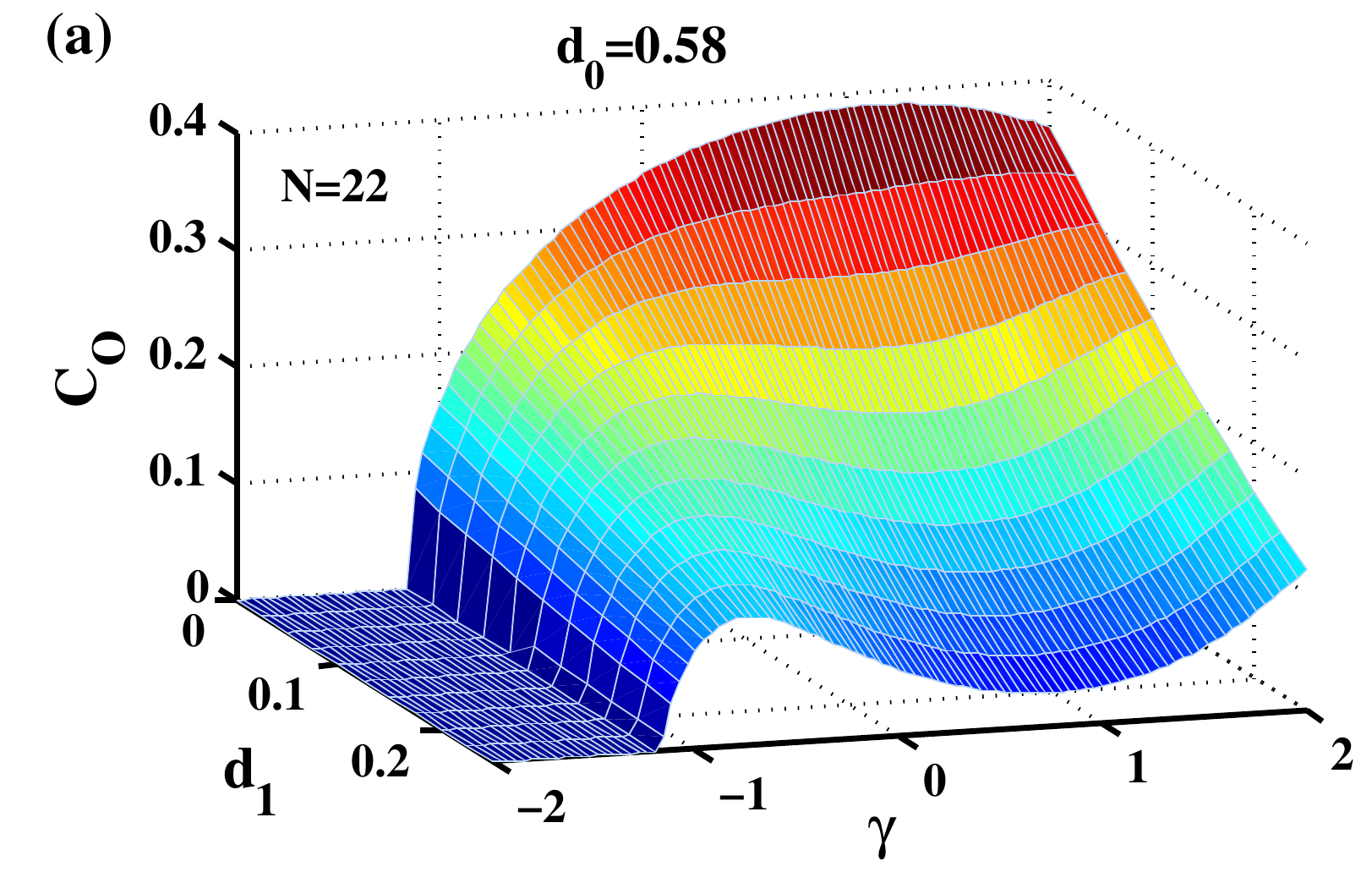,width=3.1in}} 
\centerline{\psfig{file=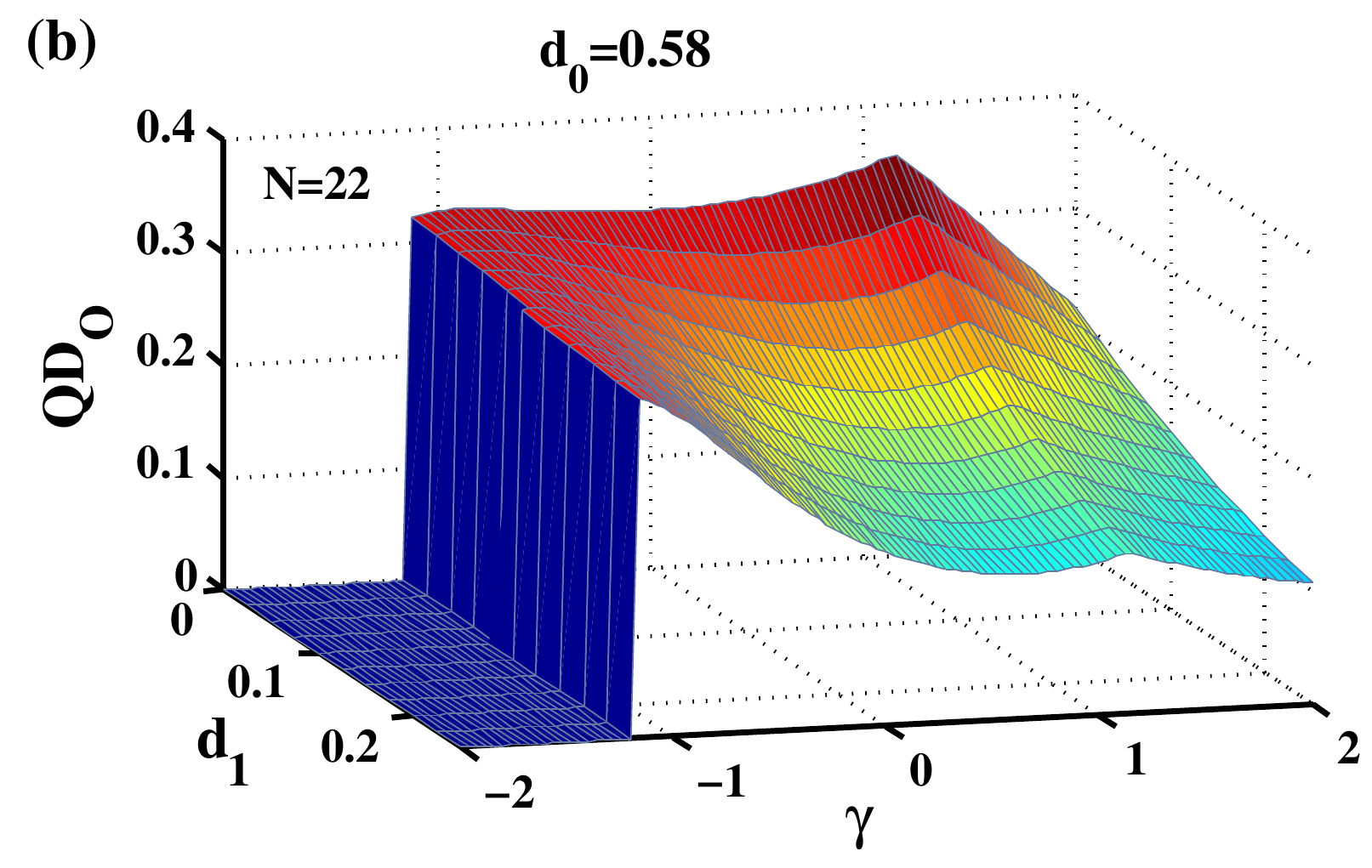,width=3.1in}} 
\caption{ (a) The concurrence and (b) the QD  between spin pairs on odd links as a function of $\gamma$ and $d_1$.  Numerical Lanczos results  for a chain size $N=22$ and $d_0=0.58$.
}
\label{C12-QD12-3D}
\end{figure}

We also investigate how the quantum correlations change for different values of the $d_1$ in a fixed $d_0$.
Fig.~\ref{C12-QD12-3D}~(a) shows $C_{O}$ as a function of $d_1$ and $\gamma$ at $d_0=0.58$.
 As it is seen $C_{O}$ has zero value in the FM phase.  For $d_1=0$, $C_{O}$ indicates two different regions correspond to FM and chiral phases.
 By increasing $d_1$, a new maximum at $\gamma_{c2}\simeq -0.7 J^{\star}$ appears that shows the BKT quantum phase
 transition at the end of the  LL phase. This maximum moves to lower values of $\gamma$ when $d_1$ increases. 
As it is already mentioned $C_{O}$ does not contain any indication which marks the critical point between composite $C1$ and $C2$ phases.
The QD between the pair spins on odd links  is depicted in panel (b) of Fig.~\ref{C12-QD12-3D}. In addition to the FM-LL phase transition, $QD_{O}$ can reveal the second boundary critical line regarding $C1-C2$ phase transition. Although, it is not able to show the critical line between LL and $C1$ phases.

\section{Summary}\label{sec4}

In this paper, we explored the finite system exact diagonalization calculations of the bipartite quantum correlation concurrence and quantum discord between nearest-neighbor pair spins to accurately identify quantum critical points in the ground state phase diagram of the spin-1/2 XXZ Heisenberg chains with alternating DM interaction. 
The Hamiltonian of the model was diagonalized by the numerical Lanczos method for different chain sizes, $N=16, 20, 22, 24$ and $26$. The existence of different phase transitions has been proven by the investigation of quantum correlations.

In the case of the pure uniform or staggered  DM interaction, the presence of the three different regions as ferromagnetic, chiral 
 and N\'{e}el phases were confirmed separately by the entanglement and the QD  with two critical points located at $\gamma_{c}=\pm\sqrt{1+d^2_{i}}$, where $i=0 (1)$ for uniform  (staggered) order.  

In the case of alternating DM interaction at $d_0d_1 \neq 0$, detection of the first critical point, as a discontinuity can be done by both entanglement and the QD, straightforwardly.
Probing the location of the maximum created in the entanglement, we found a Berezinskii-Kosterlitz-Thouless phase transition at $\gamma_{c2}\simeq -J^{\star}/\sqrt{2}$ point regarding Lattinger liquid - $C1$ phase transition.
 Moreover, we have successfully extracted the location of the BKT transition by consideration of the first derivative of the QD.
Furthermore, it was shown that the QD can distinguish another phase transition point at  $\gamma_{c3} \simeq  J^{\star}$ correspond to the quantum phase transition between two composite phases $C1$ and $C2$. 

For completeness, we have focused on the QD between pair of spins with different distances up to four lattice spacing. Results of
analyzing QD between spins with further distances show that they can reflect FM and Ising transitions as well as QD between the nearest-neighbor pair of spins.


\section{acknowledgments}
Authors acknowledge Gerardo Rossini
for helpful comments. Beradze also acknowledges support from the World Federations of Scientist Stipend Program in Georgia.



\vspace{0.3cm}

\end{document}